\documentclass[12pt]{article}
\usepackage{amsmath}
\usepackage{amssymb}
\usepackage{graphicx}
\usepackage{subfigure}
\usepackage{authblk}
\usepackage{longtable}
\begin{document}
\title{Seeing the halo rotation of nearby spiral galaxies by using {\it Planck} data}

\author[1]{Noraiz Tahir}
\author[2,3]{Francesco De Paolis}
\author[1]{Asghar Qadir}
\author[2,3]{Achille A. Nucita}
\affil[1]{School of Natural Sciences, National University of Sciences and Technology, Islamabad, Pakistan}
\affil[2]{Department of Mathematics and Physics ``Ennio De Giorgi'', University of Salento, I-73100 lecce, Italy}
\affil[3]{INFN, Sezione di Lecce, CP 193, I-73100 Lecce, Italy}
\date{\today}
\maketitle
\begin{abstract}
	The rotation of the galactic  halos is a fascinating topic which is still waiting to be addressed. {\it Planck} data has shown the existence of  a temperature  asymmetry towards  the halo of several nearby galaxies, such as M31, NGC 5128, M33, M81 and M82. However, the cause of this asymmetry is an open problem. A possibility to explain the observed effect relies on the presence of  ``cold gas clouds'' populating the galactic halos, which may be the answer to the so-called missing baryon problem. Here, we present a technique to estimate an upper limit to the rotational velocity of the halo of some nearby spiral galaxies by using both their dynamical masses and the {\it Planck} data.
\end{abstract}

%\keywords{molecular clouds, dark matter, baryons, interstellar medium, spiral galaxies}

\section{Introduction}
It is well known that baryons constitute about $5\%$ of the Universe. Moreover, about half of them are missing, and it is not known in which form they are hidden.  Different scenarios were proposed to address this issue (see, e.g., \cite{d1998,d1995,cen,fra,mat,rev}). It was proposed that molecular hydrogen clouds at the cosmic microwave background (CMB) radiation  temperature may contribute to the galactic halo dark matter \cite{d1995a}. If these clouds are at the  CMB temperature, the question is how can they be detected. A suggestion was to look for a Doppler shift in the halo of the nearby Andromeda (M31) galaxy. The part of the halo rotating away from the observer will be red-shifted and the part rotating towards the observer will be blue-shifted.  {\it Planck} data confirmed the effect for the M31 halo, and  helped to study its rotational dynamics  \cite{d2014}. After that {\it  Planck} data allowed studying the dynamics of various other galaxies and their halos \cite{d2015,g2015,d2016,g2018}. The composition, constituents and dynamics of the halo is obviously very difficult to determine. After all, by definition, this is the part of the galaxy that we cannot see but know of its presence due to the effect of its mass on the motion of the visible part. One can assume that "the halo rotates with the galaxy" but, as the galaxy does not rotate like a rigid disc, it would not be clear which part it ``rotates with''. There is need to be able to see the rotation of the halo. Due to the above effect a new window is opened, to probe the dynamics of the halos of galaxies.
 
We know the temperature asymmetry is present, but the question is what causes this effect: it might be related, to the presence of a hot, diffuse gas which emits in the CMB frequency region through Inverse Compton scattering; to synchrotron emission by fast  electrons; to anomalous microwave emission (AME) from dust grains; or to cold gas clouds populating the outer galaxy regions. Recently, a toy-model was used for the cold gas clouds in the M31 halo by using the isothermal Lane-Emden equation, provided the density at the boundary of the cloud should merge with the density of the interstellar medium \cite{n2018}. 
 
The aim of the present  paper is to estimate, under the assumption that galactic halos are populated by cold gas clouds, the rotational velocity of some nearby galaxies and by using both their dynamical masses and the temperature asymmetry towards their halos detected in  {\it Planck} data. The plan of the paper is as follows: In the next section we will use three widely adopted models for the dark matter distribution in galactic halos to describe the distribution of the cold gas clouds. Then, in section 3, we use both the galaxy dynamical mass and  {\it Planck} data to derive an estimate of the halo rotation velocity of  the NGC 5128, M33, M81, and M82 galaxies. In the last section, we offer our main conclusions. 
\section{Dark Matter Distribution in Galaxy Halos}
The three most widely adopted models to describe the dark matter distribution in galaxy halos are those proposed by Navarro, Frenk, and  White \cite{nfw}, by Moore \cite{moore} and by Burkert \cite{burk}. The corresponding density distribution are given in the  following equations 
\begin{equation}
\rho_{NFW}(r)=\frac{\rho_{c}}{\left[\frac{r}{r_{c}}\right]\left[1+\frac{r}{r_{c}}\right]^{2}},
\label{eq1}
\end{equation}
\begin{equation}
\rho_{Moore}(r)=\frac{\rho_{c}}{\left[\frac{r}{r_{c}}\right]^{1.5}\left[1+\left(\frac{r}{r_{c}}\right)^{1.5}\right]},
\label{eq2}
\end{equation}
\begin{equation}
\rho_{Burkert}(r)=\displaystyle{\frac{\rho_{c}}{\left[1+\frac{r}{r_{c}}\right]\left[1+\left(\frac{r}{r_{c}}\right)^{2}\right]}},
\label{eq3}
\end{equation} 
where, $\rho_{c}$ is the central density and $r_c$ is the so-called core radius. The values of these parameters are given in Table \ref{tab1} for the different considered galaxies, with the corresponding value of the dynamical mass \cite{CenA,M33,M81,M82}. We then assume that, as anticipated, cold gas clouds are distributed in the halos of the considered galaxies, following the mass distribution in eqs. (\ref{eq1})-(\ref{eq3}). Since the clouds are assumed to be isothermal and at the CMB temperature, we can use the Lane-Emden equation to obtain the mass density profile of each cloud. Now, a spherically symmetric cloud is described by the equation of hydrostatic equilibrium, i.e.
\begin{eqnarray}
-\frac{1}{\rho(r)}\nabla P(r)-\nabla\phi(r)=0.
\label{eq4}
\end{eqnarray}
Here, $\rho(r)$ is the density of the cloud, $P(r)$ is the pressure applied by the molecules against the gravity, and $\phi(r)$ is the gravitational potential of the system. The equation of state for an ideal isothermal gas is
\begin{equation}
P(r)=\rho(r)c_{s}^{2},
\label{eq5}
\end{equation}
where $c_{s}=(k~T_{CMB}/m_{H})^{1/2}$ is the sound speed, $k$ is the Boltzmann constant, $T_{CMB} \approx 2.7254$ K, is the CMB temperature and $m_{H} \approx 3.35 \times 10^{-25}~{\rm g}$ is the mass of single molecule of hydrogen. Now, the Poisson equation is
\begin{eqnarray}
\nabla^{2} \phi(r)=4\pi G\rho(r),
\label{eq6}
\end{eqnarray}
where, $G$ is Newton's constant. From eqs. (\ref{eq4}) and (\ref{eq5}) we have
\begin{eqnarray}
\rho(r)=\rho_{cl}exp\left(-\gamma\right),
\label{eq7}
\end{eqnarray}
Eqs. (\ref{eq6}) and (\ref{eq7}) can be rewritten into the isothermal Lane-Emden equation, which is given by
\begin{equation}
\frac{1}{\zeta^{2}}\frac{d}{d\zeta}\left(\zeta^{2}\frac{d\gamma}{d\zeta}\right)=exp(-\gamma),
\label{eq8}
\end{equation}
where,
\begin{equation}
\gamma=\frac{\phi}{c_{s}^{2}},
\label{eq9}
\end{equation}
and 
\begin{eqnarray}
\zeta=\sqrt{\frac{4\pi G\rho_{cl}}{c_{s}^{2}}}~r.
\label{eq10}
\end{eqnarray}
This procedure was first proposed independently by Bonnor \cite{bon} and Ebert \cite{eb}. Solving the Lane-Emden equation, with the boundary condition that $\gamma(\zeta=0)=0$ and $\frac{d\gamma}{d\zeta}|_{\zeta=0}=0$. The density distribution for an isothermal cloud, which is of the form 
\begin{equation}
\rho(r)=\rho_{cl}~e^{-\gamma}~.
\label{eq11}
\end{equation}
Now the total mass of the cloud can be calculated as
\begin{eqnarray}
M_{cl}=\int_{0}^{R_{cl}}4 \pi \rho(r)r^{2}~dr.
\label{eq12}
\end{eqnarray}
where, $\rho_{cl}$ is the central density of the clouds and $\gamma= \phi/c_s$. It turns out that for our cold gas clouds $c_s \simeq 3.3 \times 10^{4}~{\rm cm~s^{-1}}$. The total mass of the cloud can be obtained by integrating the mass density distribution over the cloud up to $R_{cl}$, using eq. (\ref{eq12}). Assuming that the cloud density at the boundary merges with that of the interstellar medium  (IM). The mass $M_{cl}$ and the radius $R_{cl}$ of the clouds turn out to be $M_{cl} \simeq 300  ~M_{\odot}$ and $R_{cl} \simeq 56~{\rm pc}$. We note that the mass and the radius of the clouds are independent on the gas density at the boundary. In Table \ref{tab1} we give the ratio of the central and the IM density for the galaxies i.e. the values for $\rho_c/\rho_{IM}$, where $\rho_{IM}$ is the density of the IM. In Table \ref{tab1} the mass of each cloud, and the $\rho_{IM}$, for each considered galaxy is shown. It is seen that the mass of the clouds turns out to be more or less the same for each of them.  The central mass density of the gas clouds come out to be $\rho_{cl}=8.85 \times 10^{-21}~{\rm gcm^{-3}}$.
\begin{table}[ht]
	\caption{Physical parameters of the considered nearby spiral galaxies}             % title of Table
	\label{tab1}      % is used to refer this table in the text
	\begin{center}                       % used for centering table
		\begin{tabular}{ c c c  c c c c}        % centered columns (4 columns)
			\hline\hline                 % inserts double horizontal lines
	Galaxies	&	Models & $\rho_c$ & $r_c$ & $n_{cl}/n_{IM}$ &$M_{dyn}$& $M_{cl}$   \\
		&	&  (${\rm g~cm^{-3}}$) & (${\rm kpc}$)& & ($10^{10}~M_{\odot}$) & $M_{\odot}$\\    % table heading  
			% table heading 
			\hline  
			% inserts single horizontal line			
			\\
	\textbf{Cen-A}	& NFW & $1.61 \times 10^{-25}$ &  21 & $10^{5}$ & 40& 299.5\\% inserting body of the table
			
		&	Moore& $4.40 \times 10^{-25}$ & 12& $10^{5}$ &40\\
			
		&	Burkert& $1.0 \times 10^{-24}$ & 9.8& $10^{5}$ &40 \\
		\\
	\textbf{M33}	&	NFW& $7.0 \times 10^{-25}$&  35.0& $10^{5}$ & 80& 299.5\\% inserting body of the table
	
			&	Moore& $7.64 \times 10^{-26}$  & 18.0 & $10^{5}$ & 80\\
	
			&	Burkert& $5.0 \times 10^{-25}$ & 12.0 & $10^{5}$ & 80\\
			\\
			
	\textbf{M81}	&	NFW&$5.10 \times 10^{-24}$&  10 & $10^{6}$ & 200& 298.7\\% inserting body of the table
	
		&	Moore& $7.65 \times 10^{-25}$ & 18& $10^{6}$ & 200\\
	
	    &	Burkert& $3.68 \times 10^{-24}$ & 8.9&$10^{6}$ & 200\\
	    \\
		
	\textbf{M82}	&	NFW&$4.0 \times 10^{-27}$&  11& $10^{6}$ & 0.2& 298.7\\% inserting body of the table

		&	Moore& $1.79 \times 10^{-27}$ & 13 & $10^{6}$ &0.2\\

		&	Burkert& $6.29 \times 10^{-27}$ & 9.0 & $10^{6}$ &0.2\\		
		
			\hline                                   %inserts single line
		\end{tabular}
	\end{center}   
	The central density $\rho_c$ (column 3) in units of ${\rm g~cm^{-3}}$,  the core radius $r_c$ in ${\rm kpc}$ (column 4), the ratio of the central number density of the gas cloud $n_{cl}\simeq 1 \times 10^{2} {\rm cm^{-3}}$ to the number density of the IM in the particular galaxy (column 5), the dynamical mass (column 6), and the mass of each gas cloud (column 7), are given for Centaurus-A, M33, M81, and M82 galaxies.
\end{table}

\begin{figure}%
	\centering
	\subfigure[]{%
		\label{figa}%
		\includegraphics[height=1.6in]{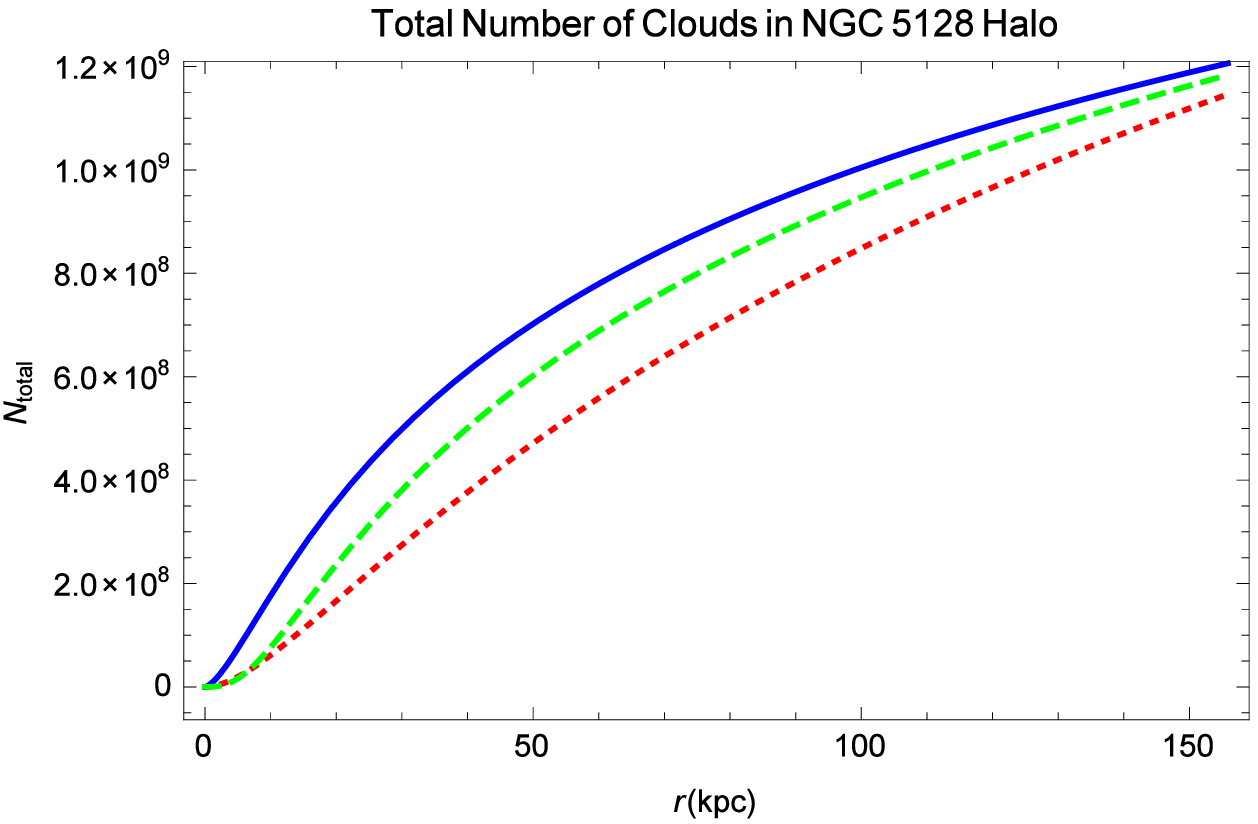}}%
	\hspace{8pt}%
	\subfigure[]{%
		\label{figb}%
		\includegraphics[height=1.6in]{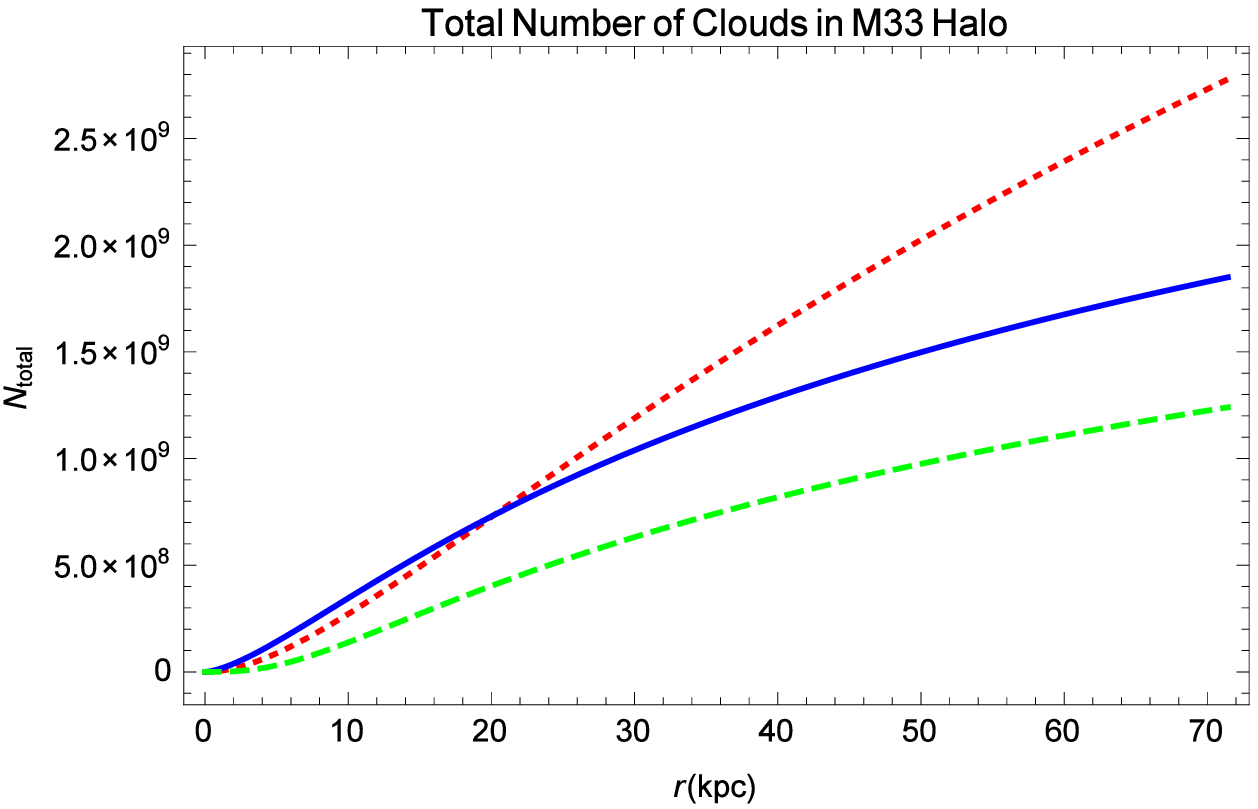}} \\
	\subfigure[]{%
		\label{figc}%
		\includegraphics[height=1.6in]{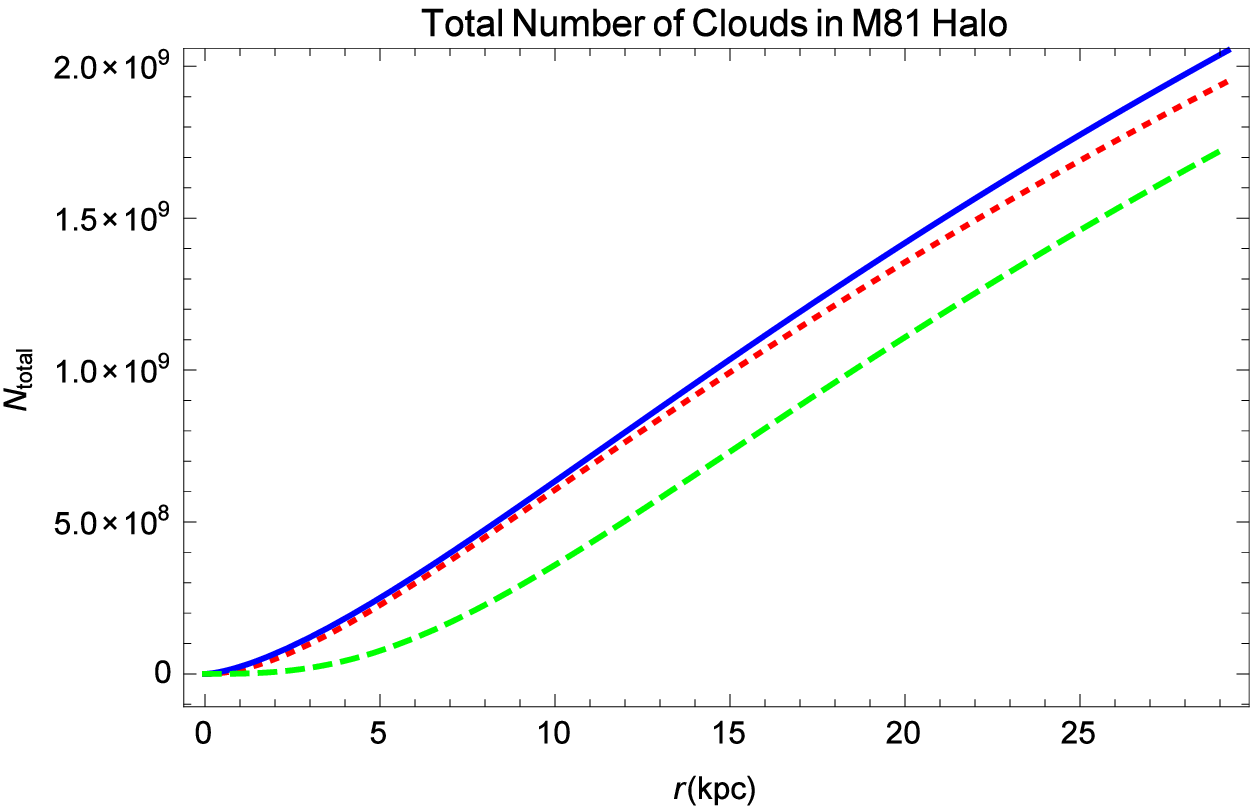}}%
	\hspace{8pt}%
	\subfigure[]{%
		\label{figd}%
		\includegraphics[height=1.6in]{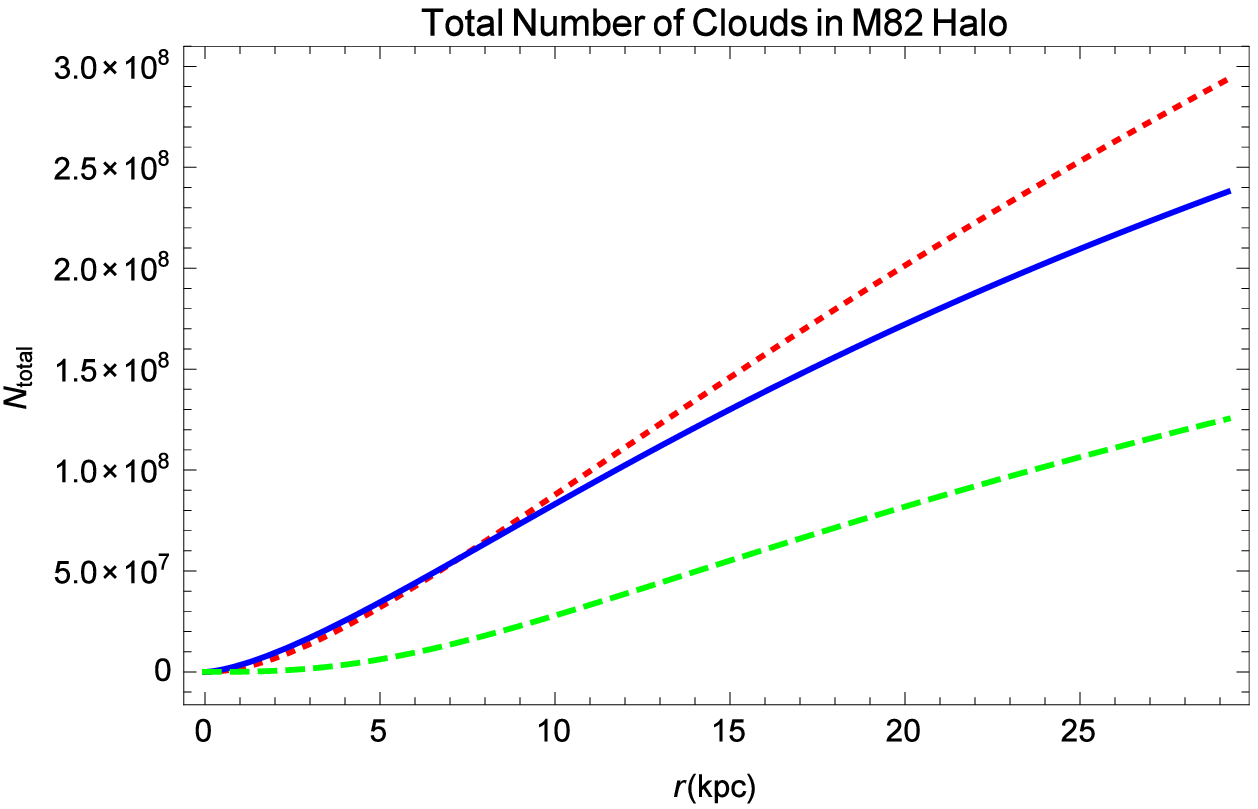}}%
	\caption{The curves in the figure represent the total number of gas clouds in the galactic halo, with the assumption that the total halo of the galaxy is composed of these clouds i.e. $f=1$. The red-dotted, blue-bold, and green-dashed, represents $N_{total}$ for NFW, Moore and Burkert models. Fig. (a) gives the total number of gas clouds in NGC 5128 within $150$ kpc; Fig. (b) gives the total number of gas clouds in M33 within $70$ kpc; Fig. (c) gives the total number of gas clouds in M81 within $25$ kpc; and Fig. (d) gives the total number of gas clouds in M82 within $25$ kpc.}%
	\label{fig1}%
\end{figure}

Once we have the mass density, mass and the radius of the gas clouds, we can obtain the total number of clouds in the galactic halos of the considered galaxies. The total number of clouds in the galactic halo is given by the relation $N_{total}=f(M_{Halo}/M_{cl})$. $M_{Halo}$ is the total mass of the halo for each galaxy, $\displaystyle{M_{Halo}(R)=\int_{0}^{R}4\pi r^{2}\rho_{N,M,B}(r)~dr}$. Here $R$ is the radius of the halo. The total number of gas clouds for each considered galaxy by assuming that the $f=1$ i.e. the full halo is composed of these clouds is shown in Fig.\ref{fig1}.      
\section{Halo Rotational Velocity}
The rotational velocity of a galactic halo can be estimated by the relation
\begin{equation}
{\rm v_{rot}}=\left(\frac{\Delta T}{T_{CMB}}\right)\left(\frac{c}{2\sin i\tau_{eff}}\right),
\label{eq13}
\end{equation}
where, $c$ is the light speed, $i$ is the inclination angle of the rotational axis of the considered galaxy  with respect to the line of sight, $\tau_{eff}=fS\overline{\tau}$ is the effective optical depth of the clouds ($f$ is the fraction of the halo dark matter in form of cold gas clouds and $S$ is the cloud filling factor  and $\overline{\tau}$ is the average optical depth of the clouds), and  $\Delta T$ is the measured  temperature asymmetry by {\it Planck} data \cite{pla} \footnote[1]{We have used the data published in the literature for the various galaxies \cite{d2015,g2015,d2016,g2018} in the 70 $GHz$, 100 $GHz$, and 143 $GHz$ bands at resolution corresponding to $N_{side}= 2048$ in HEALPix scheme \cite{hel}. The temperature excesses are shown in Table \ref{tab2} (column 4).}. The dynamical mass is defined as
\begin{eqnarray}
M_{dyn}=\frac{v_{rot}^{2}~R_{Halo}}{G},
\label{eq14}
\end{eqnarray}
where, $R_{Halo}$ is the radius of the galactic halo, and $G$ is the  Newton's gravitational constant. Eq. (\ref{eq13}) and Eq. (\ref{eq14}) yields
\begin{eqnarray}
M_{dyn}(\le R_{Halo})=\left(\frac{R_{Halo}}{G}\right)\left(\frac{\Delta T}{T_{CMB}}\frac{c}{2\sin i\tau_{eff}}\right)^{2}.
\label{eq15}
\end{eqnarray}
The values for the estimated effective optical depth $\tau_{eff}$, using Eq. (\ref{eq15}), and the corresponding halo rotational velocity ${\rm v_{rot}}$ are shown in Table \ref{tab2}.  In the case of Cen-A  ${\rm v_{rot}}$ is in the range $15-90$ $km~s^{-1}$, depending strongly on the considered {\it Planck} band but not varying much  with the value of $R_{Halo}$. In the case of the M33 galaxy, ${\rm v_{rot}}$ is between $30~km~s^{-1}$ and $150~km~s^{-1}$ and depends strongly on the considered frequency band.  For the M81 galaxy  ${\rm v_{rot}}$ turns out to be in the range  $250-450~km~s^{-1}$ while for M82 it is between $-100~km~s^{-1}$ and $345~km~s^{-1}$. In particular, in the case of M82, the halo rotation velocity (which, in our model, is assumed to be fixed and equal to that observed for the galactic disk, that is certainly a simplified assumption) seems to change sign in the outer galactic region, outside about $30-60$ kpc from the galactic center.  Actually, the  signs of a complex behavior of the M82 dynamics were already present  in the literature  and  interpreted as possibly resulting from the interaction of M82 with the M81 and/or NGC 3077 galaxies in the past \cite{g2018}.
\begin{longtable}{c c c c c c c}  
		\caption{Estimated effective optical depth and halo rotational velocity of the considered galaxies}\\
		\hline
		Galaxies & $i$ & Frequency & $R_{Halo}$ & $\Delta T$&  $\tau_{eff}$& ${\rm v_{rot}}$  \\
		& 	&  ($GHz$)  & (${\rm kpc}$) & ($\mu {\rm K}$) &($10^{-3}$) & (${\rm km~s^{-1}}$) \\   
		\hline \hline
		\endfirsthead
		\hline
		Galaxies & $i$ & Frequency & $R_{Halo}$ & $\Delta T$&  $\tau_{eff}$& ${\rm v_{rot}}$  \\
		& 	&  ($GHz$)  & (${\rm kpc}$) & ($\mu {\rm K}$) &($10^{-3}$) & (${\rm km~s^{-1}}$) \\   
		\hline \hline
		\endhead
		\hline \multicolumn{7}{r}{\textit{Continued on next page}} \\
		\endfoot
		\hline
		\endlastfoot
		\textbf{Cen-A}& $14.6^0$& 70& 92& 70& 4.80 & 89 \\
		& & & 171& 32& 4.43 & 83 \\
		& & & 245 & 12& 4.40 & 74 \\
		& & 100& 92& 60& 4.46 & 41 \\
		& & & 171& 42& 5.11 & 53 \\
		& & & 245 & 18& 5.35 & 58 \\
		& & 143& 92& 58& 3.27 & 15 \\
		& & & 171& 46& 4.00 & 23 \\
		& & & 245 & 19.5& 4.17 & 24 \\
		\textbf{M33}& $59^0$ & 70& 92& 70& 4.05 & 150 \\
		& & & 171& 30& 3.78 & 139 \\
		& & & 245 & 14& 3.71 & 125 \\
		& & 100& 92 & 60 & 3.61 & 64 \\
		& & & 171& 42& 4.28& 90\\
		& & & 245 & 18& 4.48& 98 \\
		& & 143& 92& 58.5& 2.96 & 30 \\
		& & & 171& 46& 3.25 & 38\\
		& & & 245 & 19.5& 3.49 & 41 \\ 
		\textbf{M81}& $14^0$& 70& 15 & 55.5 & 1.10 & 312 \\
		& & & 30& 65& 0.93 & 265\\
		& & & 60 & 80 & 0.99 & 253 \\
		& & 100& 15& 40& 1.66 & 366 \\
		& & & 30& 55& 1.55 & 310 \\
		& & & 60 & 75 & 1.47 & 281\\
		& & 143& 15& 45& 2.64 & 451 \\
		& & & 30& 50& 2.56 & 422 \\
		& & & 60 & 75 &  2.56 &  422\\
		\textbf{M82}& $26^0$ & 70& 15& 49.5 & 10.1 & 286 \\
		& & & 30& 42.5& 13.1 & 244  \\
		& & & 60 & -16& 11.4 & -92 \\
		& & 100& 15& 46& 9.68 & 275\\
		& & & 30& 48& 14.0 & 276 \\
		& & & 60 & -18 & 12.1 & -103\\
		& & 143& 15& 60& 11.0 & 345 \\
		& & & 30& 60 &  15.6& 346 \\
		& & & 60 & -12 & 9.9 & -69\\
		\label{tab2}
	\end{longtable}
For the considered galaxies we give the inclination angle $i$ of the rotation axis  with respect to the line of sight (column 2), the  central frequency of the considered {\it Planck} band (column 3),  the corresponding radius of the halo $R_{Halo}$ (column 4) within which the temperature asymmetry  $\Delta T$ (column 5) is detected. In the sixth column we give the estimated cloud effective optical depth $\tau_{eff}$  while in column 7  the corresponding value of the halo rotational velocity ${\rm v_{rot}}$ is shown.
\section{Conclusion}
While it is clear that spiral galaxy disks rotate, we have no idea how galactic {\it halos} rotate. To determine what the nature of the gas clouds is, we confront the model with the data, for the observed temperature asymmetry in the {\it Planck} data towards some nearby galaxies.

In spite of the oversimplified toy-model adopted to describe the gas clouds, assumed to be isothermal and at the CMB temperature, self-gravitating, entirely composed of $H_2$ molecules and rotating around the same rotation axis as the galactic disk, the obtained values of ${\rm v_{rot}}$ (see Table \ref{tab2})  appear to be  not very far from the rotation speed of the galaxy stellar disks. From Table \ref{tab1} we see that the mass of the clouds (see Table \ref{tab1} Column 7) does not change much with the change in the galaxy, it is approximately equal for NGC 5128, M33, M81, and M82. Thus, the density of the IM does not effect the mass of the cloud, as had been expected. The clouds are thought to be optically thick i.e. $\overline{\tau}=1$, This is certainly a simplifying assumption  since a correct calculation of $\overline{\tau}$ would involve summing up the integrated spectral contribution, of the individual line transitions, in a certain {\it Planck} band. The number of line transitions in the Planck spectral bands is very high, and depend on the chemical species present in each cloud. However, Kaifu et. al. \cite{kai} performed a spectral line survey toward the cold, dark cloud TMC-1 (cyanopolyyne peak) in the frequency range between 8.8 GHz and 50.0 GHz (which overlaps with the lowest frequency Planck spectral bands) by the 45 m radio telescope of Nobeyama Radio Observatory, detecting 414 lines of 38 different molecular species. These lines appear spectrally rather dense and, in the innermost region of the assumed halo, the clouds may form by the fragmentation, where the gas density is higher and the molecule collisions are more frequent, it does not appear that the clouds are nearly optically thick, while the outer regions of the clouds  should be, clearly, optically thin \cite{mw}. Since the clouds are at the CMB temperature and are neither absorbing nor emitting more radiations. It might be possible that their luminosity must be equal to the CMB luminosity. It means that their luminosity is equal to the Doppler shift observed.  Further work is needed to relax these simplifying assumptions.

We have only looked at the possibility of these cold gas clouds made of molecular hydrogen. We have not looked at the possibility of these clouds contaminated with other interstellar matter like interstellar dust. One can also try to model this possibility. These clouds might be contaminated with more or less interstellar dust or other heavier elements in the IM. but we have not discussed the contamination of radiation by matter at high temperature. That possibility needs to be eliminated, because the effect is probably negligible, but cannot be neglected without checking (Work on this issue is in progress.). One also needs to investigate how the clouds would have evolved over billions of years, from the time when the CMB temperature was higher than it is now to the present value. Precisely how did the clouds lose energy over this period? 
\section*{Acknowledgment}
We acknowledge the use of Planck's data in the Legacy Archive for Microwave Background Data Analysis (LAMBDA) and HEALPix \cite{hel} package. We would also like to acknowledge the project Inter-Asia 2018, by the collaboration of National University of Sciences and Technology, Islamabad, Pakistan and University of Salento, Lecce, Italy, and INFN which helped us to interact and work on this project. NT is grateful for the hospitality to the ISUFI college in campus, Eckotecne, of University of Salento, Lecce, Italy. FDP and AAN acknowledge the TAsP and Euclid INFN projects.

\end{document}